\documentclass[twocolumn,twocolappendix]{aastex63}

\usepackage[version=4]{mhchem}
\usepackage{amsmath,amssymb}
\received{}
\revised{}
\accepted{}
\submitjournal{ApJL}

\shorttitle{Phosphine in a prestellar core}
\shortauthors{Furuya et al.}
\newcommand{\edes}{E}

\newcommand{\ehop}{{E_{{\rm hop}}}}

\begin{document}

\title{Deep Search for Phosphine in a Prestellar Core}

\correspondingauthor{Kenji Furuya}
\email{furuya@astron.s.u-tokyo.ac.jp}

\author[0000-0002-2026-8157]{Kenji Furuya}
\affiliation{National Astronomical Observatory of Japan, Osawa 2-21-1, Mitaka, Tokyo 181-8588, Japan}
\affiliation{Department of Astronomy, Graduate School of Science, University of Tokyo, Tokyo 113-0033, Japan}
\author[0000-0002-0095-3624]{Takashi Shimonishi} 
\affiliation{Institute of Science and Technology, Niigata University, Ikarashi-ninocho 8050, Nishi-ku, Niigata 950-2181, Japan}



\begin{abstract}
Understanding in which chemical forms phosphorus exists in star- and planet-forming regions and how phosphorus is delivered to planets are of great interest from the viewpoint of the origin of life on Earth. 
Phosphine (\ce{PH3}) is thought to be a key species to understanding phosphorus chemistry, but never has been detected in star- and planet-forming regions.
We performed sensitive observations of the ortho-\ce{PH3} $1_0-0_0$ transition (266.944 GHz) toward the low-mass prestellar core L1544 with the ACA stand-alone mode of ALMA.
The line was not detected down to 3$\sigma$ levels in 0.07 km s$^{-1}$ channels of 18 mK.
The non-detection provides the upper limit to the gas-phase \ce{PH3} abundance of $5\times10^{-12}$ with respect to \ce{H2} in the central part of the core.
%%The upper limit value is four orders of magnitude lower than the elemental abundance of phosphorus in the ISM.
Based on the gas-ice astrochemical modeling, we find the scaling relationship between the gas-phase \ce{PH3} abundance and the volatile (gas and ice with larger volatility than water) P elemental abundance for given physical conditions.
This characteristic and well-constrained physical properties of L1544 allow us to constrain the upper limit to the volatile P elemental abundance of $5\times10^{-9}$, which is a factor of 60 lower than the overall P abundance in the ISM.
Then the majority of P should exist in refractory forms.
The volatile P elemental abundance of L1544 is smaller than that in the coma of comet 67P/C-G, implying that the conversion of refractory phosphorus to volatile phosphorus could have occurred along the trail from the presolar core to the protosolar disk through e.g., sputtering by accretion/outflow shocks.
\end{abstract}





\section{Introduction}
Phosphorus-bearing molecules, such as nucleic acids, 
are essential for life on Earth.
Then, understanding in which chemical forms phosphorus exist in star- and planet-forming regions and how phosphorus is delivered to planets are of great interest from the viewpoint of the origin of life on Earth.
The recent detection of phosphorus-bearing molecules, PO, in the coma of comet 67P/Churyumov-Gerasimenko \citep[][]{altwegg16,rivilla20} indicates that at least some fraction of phosphorus was present as volatiles (i.e., gas and ice) in the protosolar disk, whereas in meteorites, phosphorus is identified as inorganic minerals 
\citep{pasek05} and alkyl phosphonic acids \citep{cooper92}.

In low-mass and high-mass star-forming regions, PO and PN in the gas phase are the only neutral P-bearing molecules detected so far \citep[e.g.,][]{Tur87,Ziu87,fontani16,lefloch16,rivilla16,Min18,rivilla20,bergner22,koelemay23}.
In addition, the detection of \ce{PO+} was reported in a molecular cloud located in the Galactic center \citep{rivilla22}.
These gas-phase molecules are detected in shocked regions, but not in hot core/corino regions, implying that P in the solid phase is sputtered by shocks, and PO and PN are formed by gas-phase reactions, assisted by photochemistry \citep{jimenez-Serra18,rivilla20}.

However, to date, no infrared detection of solid-state P-bearing species are reported in any interstellar sources; hence, there is no observational constraint on solid-state P.
Astrochemical models have predicted that phosphine (\ce{PH3}) ice is the dominant carrier of volatile P in molecular clouds \citep[e.g.,][]{charnley94,chantzos20}.
Then there are several theoretical studies to study the phosphorus chemistry in the shock regions, assuming that P is initially locked up in \ce{PH3} \citep[e.g.,][]{aota12,jimenez-Serra18}.

%Gas-phase PN and PO detected in the star-forming regions are originated from the shocked regions, where \ce{PH3} ice (and/or P-bearing minerals) might be sublimated/sputtered into the gas phase, and the gas-phase chemistry could convert \ce{PH3} into PO and PN \citep[e.g.,][]{jimenez-Serra18,rivilla20,bergner22}. 
%These gas-phase molecules account for less than 1 \% of elemental P in the star-forming regions, suggesting that most P ($\sim$99 \%) is locked up in the solid phase as ices or refractory minerals or both.

Despite the possible importance of \ce{PH3} to understand the phosphorus chemistry, very little is known about \ce{PH3} in star-forming regions.
Because \ce{PH3} is relatively volatile species (the sublimation temperature is $\sim$40 K \citep{molpeceres21} and \ce{PH3} ice can be non-thermally desorbed even at $\sim$10 K, see below),
\ce{PH3} in the gas-phase may be detectable by radio observations.
However, the \ce{PH3} gas has never been detected in molecular clouds and
star-forming regions \citep[e.g.,][]{turner90,lefloch16}. 
To the best of our knowledge, the only detection of \ce{PH3} has been reported in the circumstellar envelope of a carbon star \citep{agundez14}.
%We propose ACA stand-alone observations of the \ce{PH3} $1_0-0_0$ transition (266.944 GHz) toward a prestellar core to detect \ce{PH3} for the first time.

In contrast to the poor observational understanding of \ce{PH3}, its chemistry has been studied in detail by quantum chemical calculations and laboratory experiments.
The formation of \ce{PH3} via gas-phase reactions is inefficient at low temperatures ($\sim$10 K), because of the endothermicity of relevant reactions \citep{thorne84}.
Thus, \ce{PH3} is thought to be formed on grain surfaces by the sequential hydrogenation of atomic P through barrierless reactions on grain surfaces \citep{molpeceres21}.
Once \ce{PH3} is formed on grain surfaces as ice, \ce{PH3} ice can be sublimated into the gas-phase at $\gtrsim$40 K \citep{molpeceres21} or non-thermally desorbed by chemical desorption \citep[desorption induced by the energy released by chemical reactions; ][]{nguyen20,nguyen21,furuya22a}, by photodesorption or by shock-induced desorption \citep[e.g.,][]{lefloch16,jimenez-Serra18}.
Thermal desorption should be the dominant desorption mechanism at $\gtrsim$40 K, while at lower temperatures, non-thermal desorption processes would be more efficient than thermal desorption.

Then there would be two possible approaches to detect \ce{PH3}: observing thermally desorbed \ce{PH3} in the warm ($\gtrsim$40 K) inner envelopes of protostellar sources (hot corinos/cores) or observing non-thermally desorbed \ce{PH3} in prestellar cores.
Considering the non-detection of \ce{PH3}, but the detection of PO and PN in the outflow shock region L1157-B1 \citep{lefloch16}, \ce{PH3} might be short-lived in the warm gas and quickly 
converted into PO and PN as suggested by astrochemical models \citep{aota12,jimenez-Serra18}.
Therefore, we focus on the latter approach in this work.

We report ACA stand-alone observations of the ortho-\ce{PH3} $1_0-0_0$ transition (266.944 GHz) toward the dust continuum peak of low-mass prestellar core L1544 in Taurus at a distance of 135 pc \citep{schlafly14}.
L1544 is one of the best-studied prestellar cores from both the physical and chemical points of view and is a good target for the search of \ce{PH3} \citep[e.g.,][]{crapsi07,spezzano17,furuya18b,caselli19,readelli21}.
The detection of hydrides in the gas phase toward the dust continuum peak of the core has been reported: \ce{H2O} \citep{caselli12}, \ce{NH3} \citep{caselli17}, and \ce{H2S} \citep{vastel18}.
These hydrides are likely formed on the grain surface via a sequence of hydrogenation reactions of atoms and released into the gas phase by non-thermal desorption mechanisms.
Judging from the large Einstein A coefficient and the small upper state energy \citep[12.8 K; ][]{muller05} of the ortho-\ce{PH3} $1_0-0_0$, the transition should be the strongest one in the cold gas ($<$10 K).
For example, the upper state energy of ortho-\ce{PH3} $2_0-1_0$ transition at 533.795 GHz is 38.4 K \citep[][]{muller05}, which is much higher than the gas temperature of the prestellar core.
As the critical density of ortho-\ce{PH3} $1_0-0_0$ is high ($>$10$^5$ cm$^{-3}$, Appendix \ref{sec:crit}), the central region of L1544 inside a radius of $\sim$1000 au, 
where the density is larger than 10$^6$ cm$^{-3}$ \citep{caselli19}, is the suitable place to observe the transition.
In addition, high spatial resolution observations with ACA allow us to selectively probe \ce{PH3} in the central region of the core, where \ce{PH3} is expected to be non-thermally desorbed by chemical desorption.
Gas-phase \ce{PH3} desorbed by the chemical desorption is a good probe of the elemental abundance of volatile phosphorus (see Section \ref{sec:model_result}). 

The source had been observed with the ACA stand-alone mode to map the dust continuum emission \citep{caselli19} and \ce{NH2D} lines \citep{caselli22}.
There has been no detection of P-bearing species in L1544, while \citet{turner90} and \citet{rivilla18} reported the non-detection of PN with the NRAO 12 m telescope and IRAM 30 m telescope, respectively.

%The rest of this paper is organized as follows.
%Section \ref{sec:observation} summarizes the observational details and the data reduction procedure.
%Section \ref{sec:result} presents the observational data products
%and the upper limit to the gas-phase \ce{PH3} column density.
%In Section \ref{sec:model}, the upper limit to volatile P abundance, which is available for gas-ice chemistry in the prestellar core and later evolutionary stages, is discussed based on our observations and astrochemical modeling.
%Implications of our findings are discussed in Section \ref{sec:discussion}.

\section{Observations} \label{sec:observation}
Observations were carried out with the ACA stand-alone mode of ALMA in July and August 2023 as a Cycle 9 program (2022.1.01490.S, PI: K. Furuya). 
The total number of 7 m antennas is 9-11, where the minimum–maximum baseline lengths are 8.9–48.9 m. 
The total on-source integration time is 7.3 hours. 
The pointing center of antennas is RA = 05$^\mathrm{h}$04$^\mathrm{m}$17$\fs$21 and Dec = +25$^\circ$10$\arcmin$42$\farcs$8 (International Celestial Reference System, ICRS), which corresponds to the dust continuum peak of L1544 \citep{spezzano16}. 
The sky frequencies of 250.41--250.47, 250.79--250.85, 265.86--265.92, 266.89--267.01, 267.53--267.59 GHz, are covered by five spectral windows with a velocity resolution of 0.07 km s$^{-1}$, where the forth window is used for the analysis of the PH$_3$ transition. 
The sky frequency of 252.06--253.94 GHz is also observed with a continuum mode. 
%Details of the observation settings are summarized in Table \ref{tab_Obs}. 

\begin{figure}[t!]
\plotone{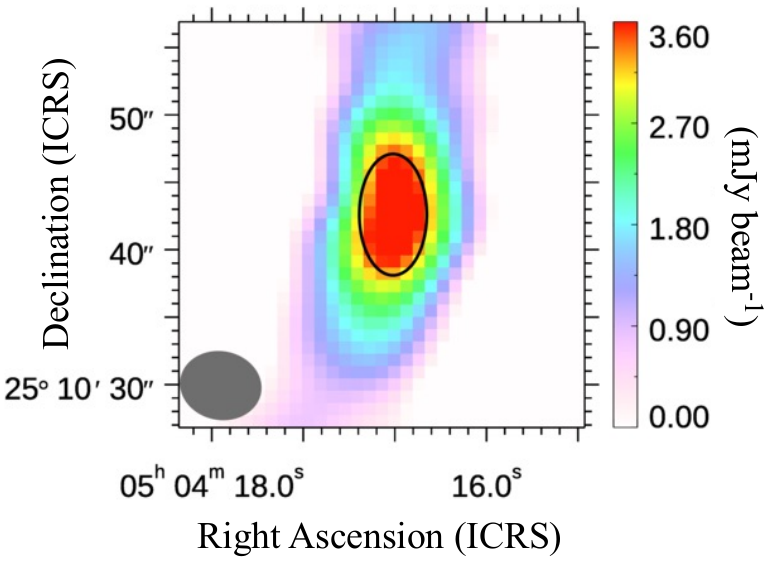}
\caption{
ALMA 1.2 mm continuum emission toward L1544. 
The black solid ellipse represents the region where we extract the spectrum and continuum flux discussed in the main text. 
The synthesized beam size is shown by the gray-filled circle. 
North is up, and east is to the left.
}
\label{fig:obs}
\end{figure}

Raw data is processed with the \textit{Common Astronomy Software Applications} (CASA) package (version 6.4.1). 
The CASA task $\mathtt{tclean}$ is used for imaging, and the masking is done using the auto-multithresh algorithm. 
The continuum emission is subtracted from the spectral data using the $\mathtt{uvcontsub}$ task in CASA. 
The synthesized beam size at the frequency of the PH$_3$ transition is 4$\farcs$8 $\times$ 5$\farcs$8 with the Briggs weighting and a robustness parameter of 0.5. 
The beam size corresponds to 650 au $\times$ 780 au at the distance of L1544. 
The maximum recoverable scale is about 25$\arcsec$. 
The synthesized image is corrected for the primary beam pattern using the $\mathtt{impbcor}$ task in CASA. 
The aggregate continuum image is constructed by using the continuum window in addition to the line-free channels selected from the other five spectral windows. 
The spectrum and continuum flux are extracted from the 5$\farcs$0 $\times$ 9$\farcs$0 
(700 au $\times$ 1260 au)
elliptical region centered at RA = 05$^\mathrm{h}$04$^\mathrm{m}$17$\fs$01 and Dec = +25$\arcdeg$10$\arcmin$42$\farcs$60, which corresponds to the continuum peak in our observations.
The elliptical region roughly corresponds to the brightest regions in the dust continuum emission (Fig. \ref{fig:obs}).

%%%%%%%%%%%
%\begin{deluxetable*}{ l c c c c c c c c c}
%\tablecaption{Observation summary \label{tab_Obs}} 
%\tablewidth{0pt} 
%\tabletypesize{\footnotesize} %preprint 
%\tablehead{
%\colhead{Proposal ID} & \colhead{Observation} & \colhead{On-source} & \colhead{Mean} & \colhead{Number} & \multicolumn{2}{c}{Baseline\tablenotemark{b}} & \colhead{} &  \colhead{} & \colhead{Channel}  \\
%\cline{6-7}  
%\colhead{ } & \colhead{Date} &  \colhead{Time} & \colhead{PWV\tablenotemark{a}} & \colhead{of} &  \colhead{Min} & \colhead{Max} & \colhead{Beam size}  & \colhead{MRS\tablenotemark{c}} &  \colhead{Spacing}  \\
%\colhead{}  & \colhead{}               &  \colhead{(min)} &  \colhead{(mm)}              & \colhead{Antennas} & \colhead{(m)} & \colhead{(m)}  & \colhead{($\arcsec$ $\times$ $\arcsec$)} & \colhead{($\arcsec$)}  &  \colhead{(kHz)}   }
%\startdata 
%2022.1.01490.S  &  2023 Jul 24 -- 2023 Aug 9  &  440.6 &  0.5--1.1 &  9-11 &  8.9 &  48.9 &  4.8 $\times$ 5.8 &   25   &  71  \\
%\enddata
%\tablenotetext{a}{Precipitable water vapor.}
%\tablenotetext{b}{Minimum and maximum baselines. }
%\tablenotetext{c}{Maximum Recoverable Scale.}
%\end{deluxetable*}
%%%%%%%%%%

\section{Observational Results} \label{sec:result}
\begin{figure}[t!]
\plotone{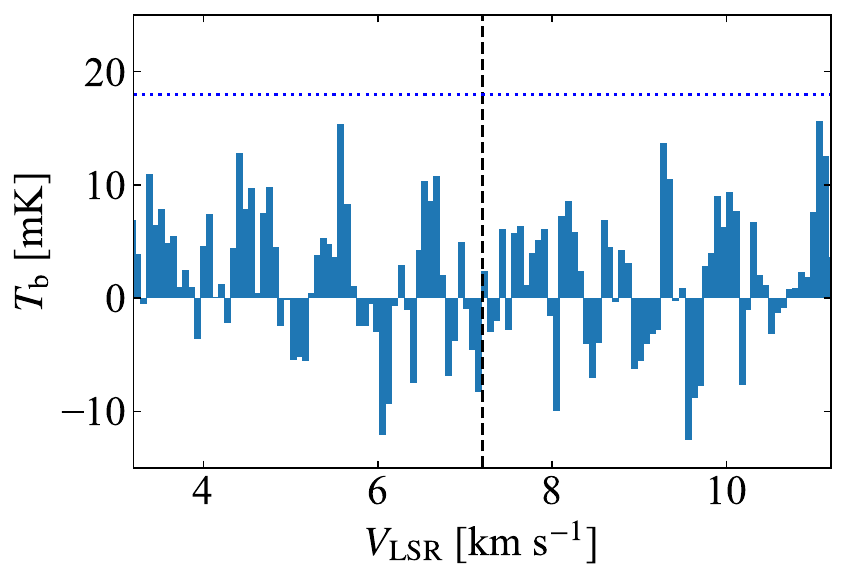}
\caption{The spectrum centered on 266.944 GHz.
The vertical line marks the $V_{\rm LSR}$ of L1544 (7.2 km s$^{-1}$).
The ortho-\ce{PH3} $1_0-0_0$ transition was not detected down to 3$\sigma$ levels in 0.07 km s$^{-1}$ channels of 18 mK (horizontal dotted line).}
\label{fig:spect}
\end{figure}

The ortho-\ce{PH3} $1_0-0_0$ transition was not detected as shown in Figure \ref{fig:spect}.
We obtain the upper limit to the \ce{PH3} column density from the 3$\sigma$ upper limit to the integrated intensity $3\sigma \sqrt{\Delta v \delta v}$, where $\sigma$ is the root-mean-square noise level of the spectra, $\Delta v$ is the FWHM line width of the spectra, and $\delta v$ is the velocity resolution.
$\Delta v$ is assumed to be 0.3 km s$^{-1}$, similar to those of the high-density tracers, \ce{H2D+} and \ce{D2H+}, obtained with IRAM 30 m single-dish telescope toward L1544 \citep{vastel04}.
Assuming local thermal equilibrium (LTE), the 3$\sigma$ intensity upper limit is converted to the upper limit to the \ce{PH3} column density ($N(\ce{PH3})$) following \citet{mangum15}.
See Appendix \ref{sec:crit} for the discussion on the validity of the LTE assumption.
The parameters of the observed transition are taken from the Cologne Database for Molecular Spectroscopy \citep{muller05} and Leiden Atomic and Molecular Database
 \citep{schoier05}.
The excitation temperature ($T_{\rm ex}$) is assumed to be the same as the gas temperature of the core center, 6 K \citep{keto15}.
As the production of \ce{PH3} by the gas-phase chemistry is negligible \citep{thorne84}, gas-phase \ce{PH3} is most likely originated from ices.
Then we assume that the ortho-to-para ratio of \ce{PH3} is unity \citep[i.e., the statistical ratio; see][]{hama16}.
As a result, the upper limit to $N(\ce{PH3})$ is estimated to be $1.2\times10^{11}$ cm$^{-2}$.
%If $T_{\rm ex}$ is assumed to be 5 K, the upper limit is $3.5\times10^{11}$ cm$^{-2}$.
Assuming the dust temperature of 6 K,
We estimate the \ce{H2} column density of the examined region to be $2.6\times10^{22}$ cm$^{-2}$ based on the dust continuum data (see Appendix \ref{sec_app_h2} for details).
The optical depth of the dust emission is estimated to be $\sim$0.001.
Accordingly, the upper limit to the gas-phase \ce{PH3} abundance is $5\times10^{-12}$ with respect to \ce{H2}. 
If $T_{\rm ex}$ is assumed to be 9 K (see Section \ref{sec:model}), the upper limits to $N(\ce{PH3})$ and the \ce{PH3} abundance are $7.6\times10^{10}$ cm$^{-2}$ and $6.7\times10^{-12}$, respectively.

As many previous observations toward L1544 were done with single-dish telescopes, IRAM 30m telescope in particular, it may be useful to provide the upper limit to $N(\ce{PH3})$ with the corresponding beam size.
We perform a similar analysis to the spectrum extracted from the 9$\farcs$2 $\times$ 9$\farcs$2 region, which corresponds to IRAM 30m’s beam at 267 GHz, centered to the continuum peak.
When $T_{\rm ex}$ = 6 K and 9 K, the upper limit to $N(\ce{PH3})$ is constrained to be $9.6\times 10^{10}$ cm$^{-2}$ and $5.9\times 10^{10}$ cm$^{-2}$, respectively.
The upper limit to the \ce{PH3} abundance is $4\times10^{-12}$ and $6\times10^{-12}$ at $T_{\rm ex} = 6$ K and 9 K, respectively.

%\epsscale{0.6}
%\epsscale{1.0}

\section{Volatile phosphorus abundance in L1544} \label{sec:model}
To constrain the volatile P abundance from the upper limit to the \ce{PH3} abundance, we conduct gas–ice
astrochemical simulations under the physical conditions appropriate for L1544.

\subsection{Model description}
We utilize the gas-ice astrochemical code based on the rate equation approach \citep[Rokko code;][]{furuya15}.
We adopt a three-phase model, where the gas phase, a surface of ice, and the chemically inert bulk ice mantle are considered \citep{hasegawa93}.
The top four ice layers are assumed to be the surface \citep{vasyunin13}, and the rest is considered as the bulk ice mantle.
Our gas-ice chemical network is based on that of \citet{garrod13} with some modification to the P chemical network as explained below.
The gas phase chemical network relevant to \ce{PH3} is updated referring to previous studies \citep{charnley94,garcia21,Fernandez-Ruz23,gomes_souza_jasper_galvao_2023, millar23}.
As the surface chemical network relevant to \ce{PH3}, we consider a sequence of hydrogenation addition reactions of atomic P to form \ce{PH3} and the hydrogen abstraction reaction from \ce{PH3} by atomic H \citep{nguyen20}.
The hydrogen addition reactions are barrierless, while the hydrogen abstraction reaction has a barrier with the tunneling corrected rate constant of $8.7 \times 10^7$ s $^{-1}$ \citep{molpeceres21}.
Laboratory experiments by \citet{turner16} showed that electron irradiation to mixed ices of  \ce{PH3} and \ce{CH4} leads to the formation of phosphanes and methylphosphanes as large as \ce{P8H10} and \ce{CH3P8H9}, whose sublimation temperature is higher than that of water.
Although it is unclear whether such a mechanism can work in the water-ice-dominated ISM ice, the current chemical network is limited to only small P-bearing molecules, such as  PN, PO, CP, and PH$_n$, where $n$ is 1-3.
The volatility of these species is higher than that of water \citep{sil21,molpeceres21}.
From this point of view, our model only considers the volatile P chemistry.

We consider a binding energy distribution for all surface species using the method developed by \citet{furuya23}, assuming the binding energy distribution follows a Gaussian distribution.
In particular, the mean binding energy of atomic H is set to be 350 K, while that of atomic P and \ce{PH_n}, where $n$ is 1-3, is taken from \citet{molpeceres21}.
Following \citet{cazaux17}, the activation energy for surface hopping from a site with the binding energy of $\edes$ to another site with the binding energy of $\edes'$ is given by
\begin{align}
\ehop (\edes \rightarrow \edes') = \chi \times {\rm min}(\edes,\,\,\edes') + {\rm max}(0,\,\,\edes - \edes'),
\end{align}
where $\chi$ is set to 0.65 for atomic H and \ce{H2} and to 0.4 for the other species.
$\ehop$ for atomic H in our model is consistent with that constrained by laboratory experiments; there are deep potential sites with $\ehop \gtrsim 350$ K, while the majority of sites are shallow sites with $\ehop \lesssim 250$ K on amorphous solid water \citep{hama12}.

Due to the low temperature in the core ($<$10 K), thermal desorption is negligible for \ce{PH_n} \citep{molpeceres21}.
As non-thermal desorption, we consider desorption by the stochastic heating by CRs \citep{hasegawa93b}, sputtering by CRs \citep[][]{dartois18}, photodesorption, and chemical desorption.
To the best of our knowledge, there is no study on the CR sputtering nor the photodesorption of P-bearing species.
We calculate the CR sputtering yield for all species in a way proposed by \citet{dartois21}, using the parameters appropriate for water ice.
We assume the yield of 10$^{-3}$ per incident UV photon for all P-bearing species.
The chemical desorption probability of \ce{PH3} upon reaction between \ce{PH2} and atomic H is set to 4 \% as constrained by laboratory experiments and their modeling \citep{nguyen21,furuya22a}.
%\citet{molpeceres23} studied the chemical desorption of PH upon the reaction between atomic P and atomic H.
Note that those studies on chemical desorption assumed that the grain surfaces are covered by water ice.
The detection of water vapor at the center of L1544 likely indicates that water ice is present on the ice surface and the water vapor is produced via photodesorption by cosmic-ray induced UV photons \citep{caselli12}.
%Based on laboratory experiments, \citet{kouchi21} have proposed that \ce{CO2} nano-crystals are embedded in the amorphous \ce{H2O} ice in intersteller ices, and a polyhedral CO crystal is attached to the \ce{H2O} ice rather than CO ice covers \ce{H2O} ice. 

There are various studies on the physical structure of L1544 based on dust continuum and molecular line observations.
As discussed below, the two of the most important physical parameters for determining the gas-phase \ce{PH3} abundance in our models are the dust temperature and the ratio of the CR ionization rate ($\xi$) to the \ce{H2} gas density ($n_{\rm H_2}$). 
%We adopt the physical structure model of \citep{keto15}.
According to the physical model of L1544 proposed by \citet{keto15}, the \ce{H2} gas density in inner $\sim$600 au of the core, where our ALMA observations prove, is $\sim$3$\times10^6$ cm$^{-3}$, while the temperature is $\sim$6-9 K.
\citet{readelli21} constrained the CR ionization rate to be $3\times10^{-17}$ s$^{-1}$ based on the observations of \ce{N2H+}, \ce{N2D+}, \ce{HC^18O+}, and \ce{DCO+} and astrochemical and radiative transfer models, adopting the physical model by \citet{keto15}.
Because the abundances of the major molecular ions depend on $\xi/n_{\rm H_2}$ \citep[][]{dalgarno06} like the gas-phase \ce{PH3} abundnance, we assume $\xi = 3\times10^{-17}$ s$^{-1}$ throughout this work.

We consider four different models, varying in temperature, gas density, and initial abundances, which are summarized in Table \ref{tab:model}.
In each model, we vary the volatile P elemental abundance from $2.57 \times 10^{-7}$ \citep[i.e., the solar value; ][]{asplund21} to  $2.57 \times 10^{-9}$, while the abundances of other elements are taken from \citet{aikawa99}.
In Models A, C, and D, initially, hydrogen is assumed to be present as \ce{H2}, while the other elements are assumed to be present as atoms or atomic ions.
In Model B, we first evolve the gas-ice chemistry for 1 Myr under the dense cloud condition ($n_{\rm H_2}=1\times10^4$ cm$^{-3}$, the temperature of 10 K, and $A_V = $10 mag), and the final abundances are used as the initial abundances of the L1544 model.

\begin{table}
\caption{Summary of the model parameters. \label{tab:model}}
\begin{center}
\begin{tabular}{cccc}
\hline\hline   
Model &  Temperature & \ce{H2} density & Initial abundance  \\
\hline
A        &  6 K & $3\times10^{6}$ cm$^{-3}$ & Atomic \\  
B        &  6 K & $3\times10^{6}$ cm$^{-3}$ & Molecular \\  
C        &  9 K & $3\times10^{6}$ cm$^{-3}$ & Atomic \\
D        &  6 K & $1\times10^{6}$ cm$^{-3}$ & Atomic \\
\hline
\end{tabular}
\end{center}
\label{table:reaction}
\end{table}

\subsection{Model results} \label{sec:model_result}
%\epsscale{0.6}
\begin{figure*}[ht!]
\plotone{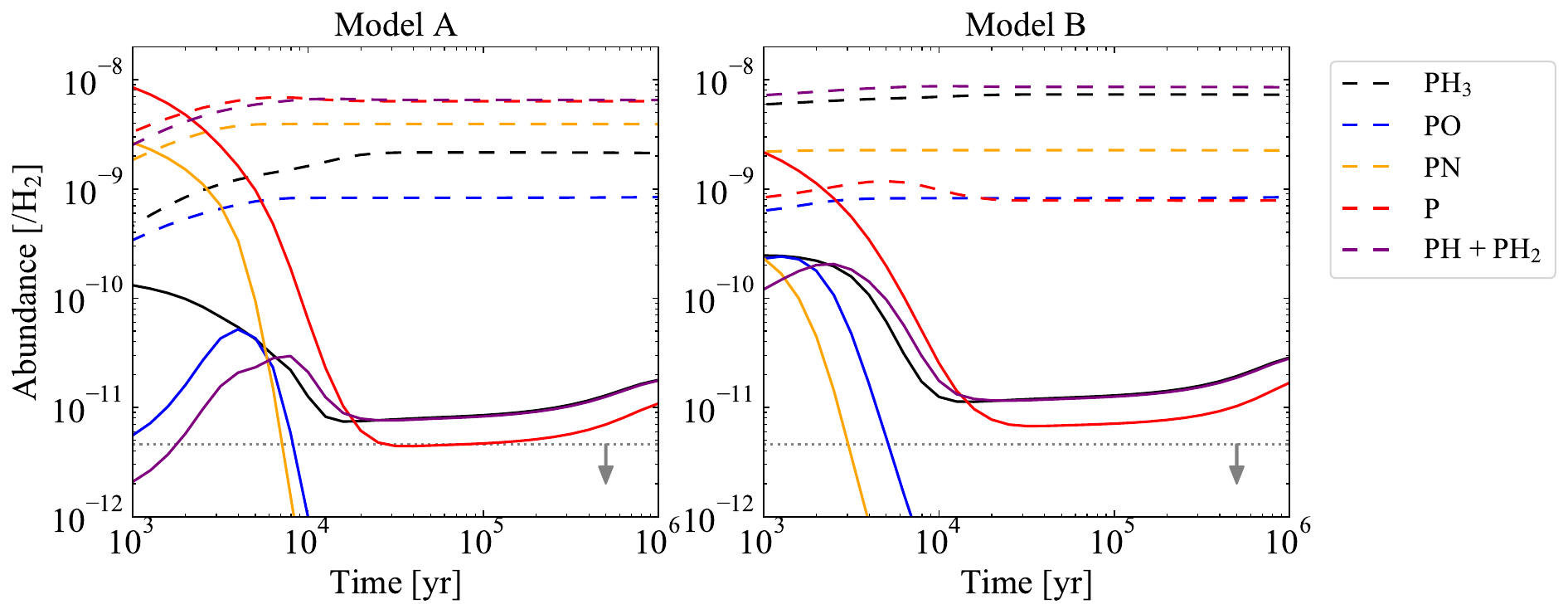}
\caption{Temporal evolution of the abundances of P-bearing species in the models with the elemental P abundance of $1\times10^{-8}$. Solid lines indicate gas-phase species, while dashed lines indicate icy species.
Purple lines represent the sum of PH and \ce{PH2} abundances to increase the visibility.
The left panel shows the results from Model A, while the right panel shows the results from Model B.
The horizontal dotted line indicates the upper limit to the gas-phase \ce{PH3} abundance in L1544.
}
\label{fig:model}
\end{figure*}
%\epsscale{1.0}

The left and right panels of Figure \ref{fig:model} show the temporal evolution of P-bearing species in Model A and Model B.
While the abundances of ice species depend on the choice of initial abundances, the gas-phase \ce{PH3} abundance is similar between the two models.  
In our models, the chemical desorption of \ce{PH3} is the dominant route for the supply of gas-phase \ce{PH3}, and the other non-thermal desorption mechanisms are negligible.
\ce{PH3} ice is formed by a sequence of hydrogenation of atomic P on grain surfaces.
Through the hydrogen abstraction reaction on the grain surface,
\begin{align}
\ce{PH3} + \ce{H} \rightarrow \ce{PH2} + \ce{H2},
\end{align}
\ce{PH3} ice is converted to \ce{PH2} ice, and the hydrogenation of \ce{PH2} ice produces \ce{PH3} ice again,
\begin{align}
\ce{PH2} + \ce{H} \rightarrow \ce{PH3}. \label{react:ph2+h}
\end{align}
When \ce{PH3} ice is produced by Reaction \ref{react:ph2+h}, 4 \% of the produced \ce{PH3} ice desorbs to the gas phase (the rest remains as ice) 
by the chemical desorption as assumed in our models.
The \ce{PH3}-\ce{PH2} loop with \ce{PH3} chemical desorption has been observed in laboratory experiments \citep{nguyen20,nguyen21}.
In our models, the \ce{PH3}-\ce{PH2} loop keeps a non-negligible fraction of \ce{PH3} in the gas phase, because \ce{PH3} can desorb in each cycle of the \ce{PH3}-\ce{PH2} loop, and because the loop continues as long as atomic H is available on grain surfaces.
The abundance of atomic H in the gas phase depends on the $\xi/n_{\rm H_2}$ ratio \citep[e.g.,][]{tielens05}, and thus the gas-phase \ce{PH3} abundance depends on the $\xi/n_{\rm H_2}$ ratio.
We also confirmed that the gas phase \ce{PH3} abundance linearly depends on the chemical desorption probability, by additionally running the modes with the desorption probability of 2 \% and 8 \%.
In our models, gas-phase \ce{PH3} is mainly destroyed by \ce{PH3} + \ce{H3+} $\rightarrow$ \ce{PH4+} + \ce{H2}, followed by the electron dissociative recombination to form \ce{PH2} + \ce{H2}.
The dissociation rate through this pathway is similar to the adsorption rate of \ce{PH3} onto dust grains. Photodissociation by CR-induced UV photons and the neutral-neutral reaction \ce{PH3} + H $\rightarrow$ \ce{PH2} + \ce{H2} are included in our models, but these are negligible compared to the pathways mentioned above.

In our models, \ce{PH3} is the most abundant gas-phase P-bearing molecule over PO and PN.
PN is mainly produced by PH + N $\rightarrow$ PN + H, while PO is mainly produced by P + OH $\rightarrow$ PO + H as in previous models \citep{charnley94,aota12,jimenez-Serra18}.
However, as the abundance of N atoms and OH in the gas phase is very low, as oxygen is locked up in \ce{H2O} ice and nitrogen is locked up in \ce{N2} and \ce{NH3} ices, a negligible fraction of atomic P and PH are converted to PO and PN.
\citet{rivilla18} reported that the PN/\ce{C^34S} column density ratio toward L1544 is lower than $\sim 8\times10^{-3}$ using the data obtained with IRAM 30m telescope \citep{jimenez16}. According to \citet{vastel18} who observed L1544 with IRAM 30m telescope, the \ce{C^34S} column density toward the dust continuum peak is $(7.8-8.3)\times 10^{11}$ cm$^{-2}$.
Taken together, the upper limit to the PN column density is $\sim6\times10^{9}$ cm$^{-2}$, which is around one order of magnitude lower than that of \ce{PH3} and corresponds to the upper limit abundance of $10^{-12}$-$10^{-13}$.
Such a low abundance of PN in L1544 is consistent with our model predictions, where the gas-phase PN abundance is $\sim$10$^{-15}$ after around 10$^4$ yr.

\begin{figure}
\plotone{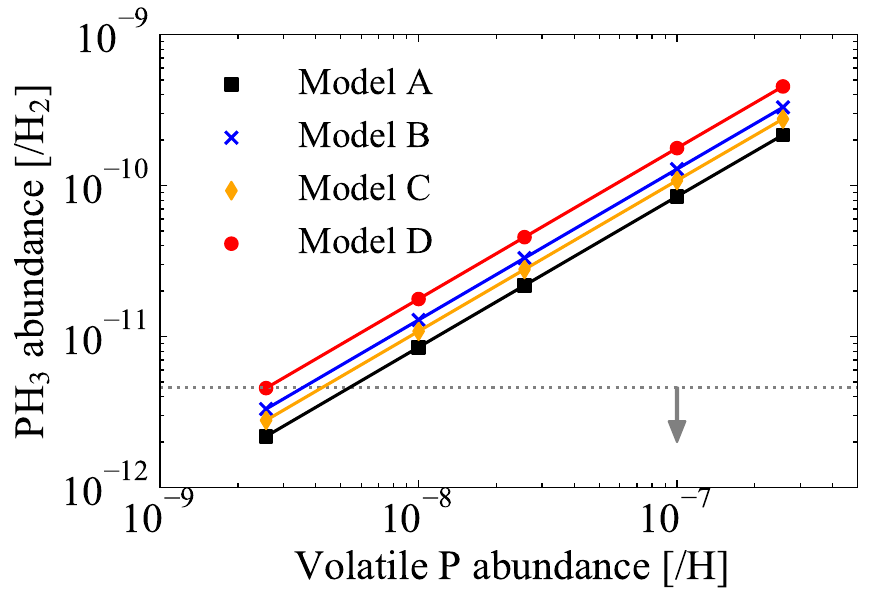}
\caption{The gas-phase \ce{PH3} abundance at 10$^5$ yr as a function of assumed P elemental abundance.
The horizontal dotted line indicates the upper limit to the gas-phase \ce{PH3} abundance in L1544.
}
\label{fig:model2}
\end{figure}

Figure \ref{fig:model2} shows the gas-phase \ce{PH3} abundance at 10$^5$ yr as a function of assumed volatile P elemental abundance.
10$^5$ yr was chosen, because at that timescale the observationally derived CO-to-\ce{H2} column density ratio of L1544 is reproduced by a gas-ice astrochemical model, considering 1D physical structure of the core \citep{vasyunin17}. 
The gas-phase \ce{PH3} abundance is almost proportional to the assumed total volatile P elemental abundance in all four models.
This characteristics allows us to constrain the volatile P abundance from the upper limit to the gas-phase \ce{PH3} abundance.
Then we conclude that the volatile P elemental abundance in L1544 is less than $5\times10^{-9}$., i.e., lower than the solar value by a factor of at least  $\sim$60.
We confirmed that this conclusion holds, even if a different time other than 10$^5$ yr ($3\times 10^4$ yr or 10$^6$ yr) is chosen.

%\epsscale{1.0}

%As the \ce{PH3} abundance scales with the elemental abundance of volatile P, we can put a constraint on the amount of volatile P in L1544 by comparing observationally derived \ce{PH3} column density with the model prediction.
%As a first step, we derive the \ce{PH3} column density using the RADEX code, assuming the density of 10$^6$ cm$^{-3}$ and the gas temperature of 7 K.
%For the collisional rate coefficients, we use the two different approaches explained above and constrain the possible column density range.

\section{Discussion and Conclusions} \label{sec:discussion}
In L1544, some hydrides have been detected.
According to previous observations with single-dish telescopes and modeling studies, the gas-phase abundances of \ce{NH3} and \ce{H2O} with respect to \ce{H2} in the central 1000 au region of L1544 are $\sim$8$\times10^{-9}$ \citep{crapsi07,caselli17} and $\sim$3$\times10^{-10}$ \citep{caselli12,keto15}, respectively.
\ce{H2S} was detected with IRAM 30m, but due to the complicated spectral shape, abundance in the central part was not constrained well \citep[See Appendix B of ][for more details]{vastel18}.
The abundances of \ce{NH3} and \ce{H2O} are more than an order of magnitude higher than the upper limit to the abundance of gas-phase \ce{PH3} constrained in this study.
Assuming the solar abundance represents the elemental abundance in the local ISM \citep[see][]{przybilla08}, the proportion of oxygen, nitrogen, and phosphorus locked up in the gas-phase hydrides in L1544 are $6\times10^{-7}$, $1\times10^{-4}$, and $<2\times10^{-5}$, respectively.
Despite nitrogen and phosphorus being in the same family, the proportion of nitrogen locked up in \ce{NH3} is much higher than the proportion of phosphorus locked up in \ce{PH3}.
This difference may partly stem from the fact that in addition to the non-thermal desorption, gas-phase \ce{NH3} can be formed via the gas-phase chemistry in the CO-depleted regions \citep[e.g.,][]{aikawa05}, while the gas phase formation of \ce{PH3} is inefficient at low temperatures ($\lesssim$10 K) \citep{thorne84}.

The solar abundance of phosphorus is $2.57\times10^{-7}$ with respect to hydrogen \citep{asplund21}.
According to the observations of UV absorption lines proving diffuse clouds, phosphorus is depleted in the gas phase, indicating that some part of phosphorus is already incorporated in refractory dust grains in diffuse clouds \citep{jenkins09,ritchey23}.
The gas phase phosphorus abundance is smaller than the solar value by a factor of up to ten, depending on sight lines \citep{ritchey23}.
\footnote{
According to \citet{ritchey23}, the transition oscillator strengths used in earlier studies \citep[e.g.,][]{lebouteiller05} to derive PII column density are in need for revision and those earlier studies overestimate the gas-phase P abundance (i.e., underestimate the degree of P depletion). More details can be found in \citet{ritchey23}.
}
%\citet{lebouteiller05} has claimed that the P/O elemental abundance ratio in diffuse cloud gas is solar, while \citet{jenkins09} has claimed that phosphorus depletion is more significant than oxygen.
%the elemental abundance of phosphorus in the gas phase is lower than the solar value by a factor of around four.
In this work, we showed that volatile (gas + ice) P abundance in prestellar core L1544 is less than $5\times10^{-9}$ and is smaller than the solar value by a factor of at least 60.
Then, almost all ($\sim$99 \% or even higher) phosphorus may be locked up in refractory dust grains during the evolution from diffuse clouds to prestellar cores.

In low-mass Class 0/I protostars, PO and PN have been detected in the outflow-shocked regions, while they are not detected in hot corinos \citep[e.g.,][]{yamaguchi11,lefloch16,bergner22}. 
This is also true for high-mass star-forming regions \citep{rivilla20,fontani24}.
%Good observational correlations are reported between P-bearing molecules and SiO, a typical strong shock tracer \citep[e.g.,][]{lefloch16}. 
These suggest that shock-induced sputtering of solid P carriers followed by the gas-phase chemistry is important for the formation of those species.
In the outflow shock region L1157-B, volatile P abundance is still a factor of $\sim$100 lower than the solar value, based on the observations of PN and PO and shock-chemical models \citep{aota12,lefloch16}, implying only a fraction of solid P is released into the gas phase.
The observational studies of PN and PO toward high-mass star-forming regions also indicate a factor of $\sim$100 lower volatile P abundance than the solar value \citep{fontani16,rivilla16}.
More observations of P-bearing molecules in prestellar cores and protostellar sources would help to understand how much P is released into the gas phase at which conditions.

In the coma of comet 67P/C-G, PO is the main reservoir of phosphorus and the P/O elemental ratio is $(0.5-2.7)\times 10^{-4}$ \citep{rubin19,rivilla20}, close to the
Solar value of $5.2 \times 10^{-4}$ \citep{asplund21}, implying that volatile P is not significantly depleted compared to L1544.
%The volatile P abundance in the coma of comet 67P/C-G seems to be larger than that in the prestellar core L1544.
Note that our astrochemical model predicts that PO is not a major reservoir of volatile P in prestellar cores (Fig. \ref{fig:model}).
Taken together, the substantial conversion of solid P in dust grains to volatile P could have occurred along the trail from the presolar core to the protosolar disk through e.g., sputtering by accretion/outflow shocks as suggested by \citet{bergner22} based on the PN and PO observations in the protostellar source.

\acknowledgments{
We are grateful to the referee for providing useful comments that helped to improve the manuscript considerably.
This work is supported in part by JSPS KAKENHI Grant numbers 20H05845, 20H05847, 21H01145, and 21K13967.
KF was supported by the ALMA Japan Research Grant of NAOJ ALMA Project, NAOJ-ALMA-324.
This paper makes use of the following ALMA data: ADS/JAO.ALMA\#2022.1.01490.S.
ALMA is a partnership of ESO (representing its member states), NSF (USA) and NINS (Japan), together with NRC (Canada), MOST and ASIAA (Taiwan), and KASI (Republic of Korea), in cooperation with the Republic of Chile. 
}

%\facility{ALMA}

%\software{CASA \citep{McM07}, Matplotlib \citep{matplotlib}}

%\begin{thebibliography}
%\end{thebibliography}






\appendix
\section{Estimation of the critical density and non-LTE analysis} \label{sec:crit} 
To check the validity of the LTE assumption, the critical density of the ortho-\ce{PH3} $1_0-0_0$ transition is estimated and the non-LTE analysis is performed. 
We assume that the main collisional partner of \ce{PH3} for excitation and de-excitation is \ce{H2}, and \ce{H2} is in the para form, because the ortho-to-para ratio of \ce{H2} is typically much less than unity in prestellar cores \citep[e.g.,][]{pagani09}.
Hereafter, let us denote ortho-\ce{PH3} as o-\ce{PH3} and para-\ce{H2} as p-\ce{H2}. Because the rate coefficients for collisions between o-\ce{PH3} and p-\ce{H2} are unavailable in the literature, we estimate them in the following two ways.
Let us denote the rate coefficients for collisions between species X and species Y as $\gamma$(X, Y).
\begin{enumerate}
\item {$\gamma$(o-\ce{NH3}, p-\ce{H2}) \citep{bouhafs17} is scaled to $\gamma$(o-\ce{PH3}, p-\ce{H2}) in a standard way, considering the mass difference between \ce{NH3} and \ce{PH3} \citep{schoier05}. This approach was adopted in  \citet{agundez14} to interpret the \ce{PH3} observation in the circumstellar envelope. Hereafter, $\gamma$ evaluated by this approach is denoted as $\gamma_1$.}
\item{Recently, $\gamma$(o-\ce{PH3}, He) were calculated by \citet{badri20}. To estimate $\gamma$(o-\ce{PH3}, p-\ce{H2}), we assume the relation, $\gamma$(o-\ce{PH3}, p-\ce{H2})/$\gamma$(o-\ce{PH3}, He) = $\gamma$(o-\ce{NH3}, p-\ce{H2})/$\gamma$(o-\ce{NH3}, He).
$\gamma$ for \ce{NH3} are taken from \citet{bouhafs17} and \citet{machin05}. Hereafter denoted as $\gamma_2$.}
\end{enumerate}
%%We find that $\gamma_2$(o-\ce{PH3}, p-\ce{H2}) is a factor of four higher than $\gamma_1$(o-\ce{PH3}, p-\ce{H2}).

The critical density of p-\ce{H2} for the transition is estimated to be $A_{10}/\gamma$, where $A_{10}$ is the Einstein A coefficient for the transition.
The critical density is $\sim$4 $\times$ 10$^5$ cm$^{-3}$ and $\sim$3 $\times$ 10$^5$ cm$^{-3}$, when $\gamma_1$ and $\gamma_2$ are adopted, respectively.
As the gas density of the central regions of L1544 ($r \lesssim 1000$ au) is $\sim$10$^6$ cm$^{-3}$ or even higher \citep{keto15,caselli19}, \ce{PH3} $1_0-0_0$ is expected to be thermalized.

To confirm the validity of the LTE assumption, we also perform non-LTE analysis with RADEX \citep{vandertak07}.
As in our LTE analysis presented in Section \ref{sec:result}, we assume $T_{\rm ex}$ of 6 K, the line width of 0.3 km/s, ortho-to-para ratio for \ce{PH3} of unity, and assume that \ce{H2} is solely present as the para-form.
When $\gamma_1$ ($\gamma_2$) is adopted, the upper limit to $N(\ce{PH3})$ is estimated to be $1.7\times10^{11}$ cm$^{-2}$ ($1.4\times10^{11}$ cm$^{-2}$) and $1.4\times10^{11}$ cm$^{-2}$ ($1.3\times10^{11}$ cm$^{-2}$) for the \ce{H2} gas density of $1\times 10^6$ cm$^{-3}$ and $3\times 10^6$ cm$^{-3}$, respectively.
These results are similar to that obtained with the LTE approximation ($<1.2\times10^{11}$ cm$^{-2}$).

While we estimated $\gamma$(o-\ce{PH3}, p-\ce{H2}) in the ways described above,
LAMBDA database provides an approximation for $\gamma$(o-\ce{PH3}, \ce{H2}) (not $\gamma$(o-\ce{PH3}, p-\ce{H2})) scaling $\gamma$(o-\ce{PH3}, \ce{He}) considering the mass difference between \ce{H2} and He.
If we use $\gamma$(o-\ce{PH3}, \ce{H2}) in the LAMBDA database as $\gamma$(o-\ce{PH3}, p-\ce{H2}), the critical density is $\sim 5 \times 10^{6}$ cm$^{-3}$, which is an order of magnitude higher than the value estimated with $\gamma_1$ or $\gamma_2$.
The difference comes from the fact that $\gamma$(o-\ce{NH3}, p-\ce{H2}) is $\sim$10 times higher than $\gamma$(o-\ce{NH3}, He) \citep{bouhafs17}.
This indicates that He is not a good proxy for \ce{H2} at least for o-\ce{NH3}.
If the collisional rate coefficient in the LAMBDA database is used, the upper limit to $N(\ce{PH3})$ is $5\times 10^{11}$ cm$^{-2}$ for the gas density of 10$^6$ cm$^{-3}$ and $2\times 10^{11}$ cm$^{-2}$ for the gas density of $3\times 10^6$ cm$^{-3}$. These values are a factor of a few larger than the value obtained with the LTE approximation.
The collisional rate coefficient for the \ce{PH3}-\ce{H2} system is highly desired.

%%Taken the o-\ce{PH3} column density of $1\times10^{12}$ cm$^{-3}$ and the gas temperature of $\sim$7 K from the models with the P depletion factor of 100, and assuming that the \ce{H2} density is 10$^{6}$ cm$^{-3}$ (the conservative value) and the velocity width of 0.5 km/s (Vastel et al. 2006), the expected intensity of \ce{PH3} $1_0-0_0$ toward the core center 
%is $\sim$70 mK (when $\gamma_1$ is adopted) or $\sim$90 mK ($\gamma_2$).
%Our requested sensitivity of 14 mK enables us to detect \ce{PH3} $1_0-0_0$ with SNR $\gtrsim$5. 

\section{Derivation of the H$_2$ column density} \label{sec_app_h2} 
Based on the standard treatment of optically thin dust emission, we apply the following equation to calculate the H$_2$ column density ($N({\ce{H2}})$):
\begin{equation}
N({\ce{H2}}) = \frac{F_{\nu} / \Omega}{2 \kappa_{\nu} B_{\nu}(T_{d}) Z \mu m_{\mathrm{H}}} \label{Eq_h2}, 
\end{equation}
where $F_{\nu}/\Omega$ is the continuum flux density per beam solid angle as estimated from the observations, $\kappa_{\nu}$ is the mass absorption coefficient of dust grains coated by thick ice mantles taken from \citet{Oss94} and we use 1.21 cm$^2$ g$^{-1}$ at 1.16 mm, $T_{d}$ is the dust temperature and we assume the gas and dust temperatures are the same in this work, $B_{\nu}(T_{d})$ is the Planck function, $Z$ is the dust-to-gas mass ratio and we use a canonical value of 0.008 for the solar neighborhood, $\mu$ is the mean atomic mass per hydrogen \citep[1.41,][]{Cox00}, and $m_{\mathrm{H}}$ is the hydrogen mass. 

Our $N({\ce{H2}})$ is a factor of $\sim$4 lower than the value obtained by \citet{crapsi05} based on IRAM 30m observations at 1.2 mm ($2.6 \times 10^{22}$ cm$^{-2}$ versus $\sim 1 \times 10^{23}$ cm$^{-2}$).
The dust opacity adopted in \citet{crapsi05} is a factor of $\sim$2 larger than our adopted value.
On the other hand, \citet{crapsi05} assumed the dust temperature of 10 K, which is higher than our assumed value of 6 K. As $B_{\nu}(10\,{\rm K})/B_{\nu}(6\,{\rm K}) \sim$ 1.7 at 1.2 mm, the difference due to the dust opacity and the dust temperature is almost canceled out.
%We are not fully sure about the cause of the difference.
The contribution from the largely-extended ($>$25") continuum emission that was resolved out in the ACA observations may account for the above discrepancy. 
%Even if our $N({\ce{H2}})$ is underestimated, we overestimate the upper limit to the \ce{PH3} abundance (i.e., underestimate the depletion factor of volatile P).

%\section{Potentail impact of Missing flux}
%{\bf
%Besides the \ce{PH3} line, the emission lines of HCO+(3-2) and HCN(3-2) are covered in this work, and they are detected. However, these molecular species are often largely extended and have a different chemical origin (gas-phase production) from PH3. Actually, the comparison of the present ACA flux of HCO+(3-2) with those of CSO or JCMT reported in \citet{gregersen00} or \citet{schnee13} suggests that only less than 1 \% of the total flux is recovered by ACA for largely extended HCO+(3-2) emission. 

%On the other hand, the figure below from \citet{caselli22} shows the \ce{NH2D} line (110.15355 GHz) observed towards the L1544 core center. The black line is obtained by the ALMA 12m+ACA for the 22” diameter region centered at the L1544 core, while the red one is obtained by IRAM 30-m for the same region. The IRAM 30-m flux of the NH2D line is almost recovered by the ACA data, as claimed by \citet{caselli22}. We presume that this is also the case for the PH3 line, since both lines have a relatively high critical density (1e5 cm-3 for the \ce{NH2D} line and even higher for the \ce{PH3} line) and both species have a similar chemical origin (i.e., related with grain surface chemistry). Taking into account that PH3 is chemically more similar to \ce{NH2D} than \ce{HCO+}, we expect that the missing flux of the \ce{PH3} line is not significant as in the case of \ce{NH2D}. 
%}

\bibliography{ms}{}
\bibliographystyle{aasjournal}

\end{document}